\documentstyle[12pt]{article}
\newlength{\extraspace}
\newlength{\extraspaces}
\newcommand{\be}{\begin{equation}
\addtolength{\abovedisplayskip}{\extraspaces}
\addtolength{\belowdisplayskip}{\extraspaces}
\addtolength{\abovedisplayshortskip}{\extraspace}
\addtolength{\belowdisplayshortskip}{\extraspace}}
\newcommand{\ee}{\end{equation}}

\newcommand{\ba}{\begin{eqnarray}
\addtolength{\abovedisplayskip}{\extraspaces}
\addtolength{\belowdisplayskip}{\extraspaces}
\addtolength{\abovedisplayshortskip}{\extraspace}
\addtolength{\belowdisplayshortskip}{\extraspace}}
\newcommand{\ea}{\end{eqnarray}}

\newcommand{\nonu}{\nonumber \\[.5mm]}
\newcommand{\A}{&\!\!\!}

\newcommand{\il}{{\lambda_{{}_{{}_{\!\!\!\!\scriptstyle{i}}}}}{^{}{^{}{^{}}}}_{\mu}}

\newcommand{\newsection}[1]{
\vspace{7mm} \pagebreak[3] \addtocounter{section}{1}
\setcounter{subsection}{0} \setcounter{footnote}{0}
\begin{center}
{\large {\bf \thesection. #1}}
\end{center}
\nopagebreak
\medskip
\nopagebreak \hspace{3mm}}

\begin{document}

\begin{center}
{{\bf A Kerr Metric Solution in Tetrad Theory of Gravitation}}
\end{center}
\centerline{ Gamal G.L. Nashed}

\bigskip

\centerline{{\it Mathematics Department, Faculty of Science, Ain
Shams University, Cairo, Egypt }}

\bigskip
 \centerline{ e-mail:nashed@asunet.shams.eun.eg}

\hspace{2cm}
\\
\\
\\
\\
\\
\\
\\
\\

Using an axial parallel vector field we obtain two  exact
solutions of a vacuum gravitational field equations. One of the
exact solutions gives the Schwarzschild metric while  the other
gives the Kerr metric. The parallel vector field of the Kerr
solution have an axial symmetry. The exact solution of the Kerr
metric contains two constants of integration, one being the
gravitational mass of the source and the other constant $h$ is
related to the angular momentum of the rotating source, when the
spin density ${S_{i j}}^\mu$ of the gravitational source
satisfies\\ $\partial_\mu {S_{i j}}^\mu=0$. The singularity of the
Kerr solution is studied.

\newpage
\begin{center}
\newsection{\bf Introduction}
\end{center}

The Riemann Cartan space-time is characterized by the non
vanishing of the curvature tensor and the torsion tensor
\cite{HS}. When we use the teleparallel condition the space-time
reduces to the Weitzenb{\rm $\ddot{o}$}ck space-time which is
characterized by the torsion tensor only \cite{Wr}. There are many
theories constructed using the Weitzenb{\rm $\ddot{o}$}ck
space-time \cite{HS,MW,Mc}. Among these theories is the theory
given by Hayashi and Shirafuji which is called new general
relativity and  invariant under global Lorentz transformations but
not under local Lorentz transformations \cite{HS}.

This theory contains three dimensionless parameters $a_1$, $a_2$
and $a_3$. Two of these parameters were determined by comparison
with solar-system experiments, while an upper bound was estimated
for the $a_3.$ It was found that the numerical value of $(a_1
+a_2)$ should be very small, consistent with being zero.
Throughout this paper we will assume this condition.

S$\acute{e}$az \cite{Sd} obtained a set of tetrads satisfying M\o
ller field equations and leading to the Kerr metric by using a
particular procedure but without giving an explicit form of this
tetrads. Fukui and Hayashi \cite{FH} has pointed out that axially
symmetric and stationary solution can be obtained in the new
general relativity but also without giving an  explicit form. Toma
\cite{Tn} gives an exact solution to the vacuum field equation of
the new general relativity.

It is our aim to obtain an exact axially symmetric solution of the
gravitational field equation of the new general relativity using
another procedure different from that used by Toma. In section 2
we briefly review the new general relativity. In section 3 we
apply the parallel vector field which is axially symmetric to the
field equation of the new general relativity. Two particular exact
solutions of the resulting differential equations are obtained.
One of these solutions contains one constant of integration and
gives the Schwarzschild metric. The other contains two constants
of integration and  gives the Kerr metric. The physical meaning of
these two constants is discussed in section 4. The singularity of
the Kerr solution is discussed in section 5. The final section is
devoted to the main results.

\newsection{The teleparallel theory of gravitation}

The fundamental fields of gravitation are the parallel vector
fields ${b_k}^\mu$. The component of the metric tensor $g_{\mu
\nu}$ are related to the dual components ${b^k}_\mu$ of the
parallel vector fields by the relation \be g_{\mu \nu}=\eta_{i j}
{b^i}_\mu{b^j}_\nu, \ee
 where $\eta_{i j}=diag.(-,+,+,+)$. The nonsymmetric connection
 ${\Gamma^\lambda}_{\mu \nu}$\footnote{Latin indices $(i,j,k,\cdots)$ designate the vector
number, which runs from $(0)$ to
 $(3)$, while Greek indices $(\mu,\nu,\rho, \cdots)$ designate the world-vector components
running from 0 to 3. The spatial part of Latin indices is denoted
by $(a,b,c,\cdots)$, while that of Greek indices by $(\alpha,
\beta,\gamma,\cdots)$.} are defined by
 \be
{\Gamma^\lambda}_{\mu \nu} ={b_k}^\lambda {b^k}_{\mu,\nu}, \ee as
a result of the absolute parallelism \cite{HS}.

The gravitational Lagrangian ${\it L}$ of this theory is an
invariant constructed from the quadratic terms of the torsion
tensor
 \be
 {T^\lambda}_{\mu \nu} \stackrel{\rm def.}{=}{\Gamma^\lambda}_{\mu
 \nu}-{\Gamma^\lambda}_{\nu \mu}. \ee
  The following Lagrangian
\be {\cal L} \stackrel{\rm def.}{=} -{1 \over 3\kappa} \left(
t^{\mu \nu \lambda} t_{\mu \nu \lambda}- v^\mu v_\mu \right)+\zeta
a^\mu a_\mu, \ee is quite favorable experimentally \cite{HS}. Here
$\zeta$ is a constant parameter,  $\kappa$ is the Einstein
gravitational constant and $t_{\mu \nu \lambda}, v_\mu$ and
$a_\mu$ are the irreducible components of the torsion tensor:

\ba t_{\lambda \mu \nu}\A=\A {1 \over 2} \left(T_{\lambda \mu
\nu}+T_{\mu \lambda \nu} \right) +{1 \over 6} \left( g_{\nu
\lambda}V_\mu+g_{\mu \nu}V_\lambda \right) -{1 \over 3} g_{\lambda
\mu}V_\nu,\nonu
V_\mu \A=\A {T^\lambda}_{\lambda \nu}, \nonu
a_\mu \A=\A {1 \over 6} \epsilon_{\mu \nu \rho \sigma}T^{\nu \rho
\sigma},
 \ea
where $\epsilon_{\mu \nu \rho \sigma}$ is defined by \be
\epsilon_{\mu \nu \rho \sigma} \stackrel{\rm def.}{=} \sqrt{-g}
 \delta_{\mu \nu \rho \sigma} \ee with $\delta_{\mu \nu \rho
\sigma}$ being completely antisymmetric and normalized as
$\delta_{0123}=-1$.

By applying the variational principle to the Lagrangian (4), the
gravitational field equation are given by \cite{HS}\footnote{We
will denote the symmetric part by ( \ ), for example, $A_{(\mu
\nu)}=(1/2)( A_{\mu \nu}+A_{\nu \mu})$ and the  antisymmetric part
by the square bracket [\ ], $A_{[\mu \nu]}=(1/2)( A_{\mu
\nu}-A_{\nu \mu})$ .}: \be G_{\mu \nu} +K_{\mu \nu} = -{\kappa}
T_{(\mu \nu)}, \ee \be {b^i}_\mu{b^j}_\nu
\partial_\lambda(\sqrt{-g} {J_{i j}}^{\lambda})=\lambda
\sqrt{-g}T_{[\mu \nu]}, \ee where the Einstein tensor $G_{\mu
\nu}$ is defined by \be G_{\mu \nu}=R_{\mu \nu}-{1 \over 2} g_{\mu
\nu} R, \ee \be {R^\rho}_{\sigma \mu \nu}=\partial_\mu \left
\{_{\sigma \nu}^\rho \right\}-\partial_\nu \left \{_{\sigma
\mu}^\rho \right\}+\left \{_{\tau \mu}^\rho \right\} \left
\{_{\sigma \nu}^\tau \right\}-\left \{_{\sigma \mu}^\tau \right\}
\left \{_{\tau \nu}^\rho \right\}, \ee \be R_{\mu
\nu}={R^\rho}_{\mu \rho \nu}, \ee \be R=g^{\mu \nu}R_{\mu \nu},
\ee and $T_{\mu \nu}$ is the energy-momentum tensor of a source
field of the Lagrangian $L_m$ \be T^{\mu \nu} = {1 \over
\sqrt{-g}} {\delta {\cal L}_M \over \delta {b^k}_\nu} b^{k \mu}
\ee with ${L_M}= {{\cal L}_M/\sqrt{-g}}$. The tensors $K_{\mu
\nu}$ and $J_{i j \mu}$ are defined as \be K_{\mu \nu}={\kappa
\over \lambda}\left( {1 \over 2} \left[{\epsilon_\mu}^{\rho \sigma
\lambda}(T_{\nu \rho \sigma}-T_{\rho \sigma
\nu})+{\epsilon_\nu}^{\rho \sigma \lambda}(T_{\mu \rho
\sigma}-T_{\rho \sigma \mu}) \right]a_\lambda-{3 \over 2} a_\mu
a_\nu-{3 \over 4}g_{\mu \nu} a^\lambda a_\lambda \right), \ee \be
J_{i j \mu}=-{3 \over 2} {b_i}^\rho {b_j}^\sigma \epsilon_{\rho
\sigma \mu \nu} a^\nu, \ee respectively. The dimensionless
parameter $\lambda$ is defined by \be {1 \over \lambda}={4 \over
9} \zeta+{1 \over 3 \kappa}.\ee In this paper we are going to
consider the vacuum gravitational field: \be T_{(\mu \nu)}=T_{[\mu
\nu]}=0.\ee
\newsection{ Axially symmetric solution}

The covariant form of a tetrad space having  axial symmetry in
spherical polar coordinates, can be written as \cite{Sd}

 \be
\left( {b^i}_\mu \right)= \left( \matrix{ \displaystyle{i \over A}
& 0 & 0 & iB \vspace{3mm} \cr 0 &  C \cos\phi & 0 & -r F \sin\phi
\vspace{3mm} \cr 0 & 0 & E & 0 \vspace{3mm} \cr 0 & C \sin\phi & 0
& r F \cos\phi \cr } \right), \ee  where  $A$, $B$, $C$, $E$ and
$F$ are five unknown functions of  $(r)$ and $(\theta)$ and he
zeroth vector ${b^i}_\mu$ has the factor $i=\sqrt{-1}$ to preserve
Lorentz signature. We consider an asymptotically flat space-time
in this paper, and impose the boundary condition that for $r
\rightarrow \infty$
 the tetrad (18) approaches the tetrad of Minkowski space-time,
$\left(\il \right)= {\rm diag}(i,{\delta_a}^{\alpha})$.

Applying (18) to the field equations (7) and (8), we get \ba \A \A
{-1 \over 2 r^2A^2 E^3F^2}\Biggl(B^2 E^2{A_r}^2+2AB E^3 A_rB_r
+2r^2A^2 E^2FE_rF_r+2rA^2 E^2F^2E_r-2r^2AE^2F^2A_r E_r \nonu
\A \A +2r^2A^2C^2EFF_{\theta \theta}-2r^2A^2C^2FE_\theta
F_\theta-2ABC^2EA_\theta B_\theta +2r^2AC^2F^2A_\theta
E_\theta+4r^2C^2EF^2{A_\theta}^2 \nonu
\A \A -2r^2AC^2EF^2A_{\theta
\theta}-2rAE^3F^2A_\theta-2r^2AC^2EFA_\theta F_\theta-2r^2AE^3FA_r
F_r+A^2E^3{B_r}^2\nonu
\A \A -B^2C^2E{A_\theta}^2-A^2C^2E{B_\theta}^2 \Biggr)=0,
 \ea
 \ba
\A \A {1 \over r^2A^2CEF^2} \Biggl (ABCEA_\theta B_r+B^2CEA_r
A_\theta+ABCEA_r B_\theta-2r^2CEF^2A_r
A_\theta-rA^2CEFF_\theta\nonu
\A \A -r^2A^2CEFF_{r \theta}+A^2CEB_r B_\theta+r^2A^2EFC_\theta
F_r+rA^2EF^2C_\theta+r^2A^2CFE_r F_\theta\nonu
\A \A -r^2AEF^2A_r C_\theta-r^2ACF^2A_\theta
E_r+r^2ACEF^2A_{r\theta} \Biggr)=0, \ea \ba \A \A {2 \over
r^2A^2C^3F^2} \Biggl(2rACE^2F^2A_r+2r^2AC^3FA_\theta
F_\theta+2r^2ACE^2FA_r F_r+2r^2ACE^2F^2A_{rr}+2AC^2F^2A_\theta
C_\theta\nonu
\A \A-4r^2CE^2F^2{A_r}^2+2r^2A^2E^2FC_r
F_r+2rA^2E^2F^2C_r-2r^2A^2C^2FC_\theta F_\theta+2ABCE^2A_r B_r
\nonu
\A \A -2r^2A^2CE^2FF_{rr} -4rA^2CE^2FF_r-2r^2AE^2F^2A_r
C_r-2ABC^3A_\theta B_\theta-A^2C^3{B_\theta}^2\nonu
 \A \A
-B^2C^3{A_\theta}^2+A^2CE^2{B_r}^2+B^2CE^2{A_r}^2 \Biggl)=0, \ea
\ba \A \A {-1 \over 2r^2A^2C^3E^3F^2}
\Biggl(3B^4CE^3{A_r}^2+r^2A^2CE^3F^2{B_r}^2+r^2A^2C^3EF^2{B_\theta}^2-
2r^2A^2BE^3F^2B_r C_r \nonu
\A \A -2r^2A^2BC^3F^2B_\theta
E_\theta+3A^2B^2C^3E{B_\theta}^2-4rA^2BCE^3F^2B_r+3A^2B^2CE^3{B_r}^2-
7r^2B^2C^3EF^2{A_\theta}^2 \nonu
\A \A +2r^2A^2BC^2EF^2B_\theta C_\theta+2r^2A^2BCE^2F^2B_r
E_r-4r^2A^2BCE^3FB_r F_r+6AB^3CE^3A_r B_r \nonu
\A \A-4r^2AB^2CE^3FA_r F_r-4r^2A^2BC^3EF B_\theta
F_\theta-4r^2AB^2C^3EFA_\theta F_\theta-7r^2B^2CE^3F^2{A_r}^2
\nonu
\A \A +6AB^3C^3E A_\theta B_\theta-4rAB^2CE^3F^2A_r
+2rA^2BC^2E^3FB_r+2rAB^2C^2E^3FA_r-2r^4AC^3EF^4 A_{\theta \theta}
\nonu
\A \A +2r^2A^2BC^3EF^2B_{\theta \theta} +2r^2A^2BCE^3F^2B_{r
r}+2r^4AC^3F^4A_\theta E_\theta-2r^2ABC^3EF^2A_\theta
B_\theta+2r^4AE^3F^4A_r C_r\nonu
\A \A -2r^2ABCE^3F^2A_r B_r-2r^4AC^2EF^4A_\theta C_\theta
+2r^2AB^2C^2EF^2A_\theta
C_\theta+2rA^2B^2E^3F^2C_r-2r^4A^2E^2F^4C_r E_r\nonu
\A \A -2r^4A^2C^2F^4C_\theta E_\theta-2r^4ACE^2F^4A_r
E_r+4r^4C^3EF^4{A_\theta}^2
+4r^4CE^3F^4{A_r}^2+2r^4A^2CE^2F^4E_{rr} \nonu
\A \A - 2A^2B^2C^3EFF_{\theta
\theta}-2r^2A^2B^2CE^3FF_{rr}-2r^2A^2B^2CE^2F^2E_{rr}-
2r^2A^2B^2C^2EF^2C_{\theta \theta}+2r^2AB^2C^3EF^2A_{\theta
\theta} \nonu
\A \A +2r^2AB^2CE^3F^2A_{rr}+2r^4A^2C^2EF^4C_{\theta
\theta}-2r^4ACE^3F^4A_{rr}+2r^2A^2B^2C^3FE_\theta
F_\theta+2r^2A^2B^2C^2F^2C_\theta E_\theta \nonu
\A \A -4rA^2B^2CE^3F_r-2rA^2B^2CE^2F^2E_r+2r^2A^2B^2E^3FC_r
F_r+2r^2A^2B^2E^2F^2C_rE_r-2r^2AB^2E^3F^2A_r C_r \nonu
 \A \A +2r^2AB^2CE^2F^2A_r E_r -2r^2AB^2C^3F^2A_\theta E_\theta
 -2r^2A^2B^2C^2EFC_\theta
F_\theta \nonu
\A \A -2r^2A^2B^2CE^2FE_rF_r+3B^4C^3E{A_\theta}^2 \Biggr)=0, \ea
\ba
 \A \A {1 \over 2r^2A^3C^3E^3F^2} \Biggl(
2r^2A^2BCE^3FF_{rr}+2r^2A^2BC^2EF^2C_{\theta
\theta}+2r^2A^2BCE^2F^2E_{rr}\nonu
\A \A +4r^2BCE^3F^2{A_r}^2-r^2A^2C^3F^2B_{\theta
\theta}+2r^2A^2BC^3EFF_{\theta \theta}-r^2ABC^3EF^2A_{\theta
\theta}-r^2A^2CE^3F^2B_{rr} \nonu
\A \A -r^2ABCE^3F^2A_{rr}+r^2A^2E^3F^2B_r
C_r+2rA^2CE^3F^2B_r+r^2ABE^3F^2A_r C_r\nonu
\A \A +r^2ABC^3F^2A_\theta E_\theta+2r^2ACE^3F^2A_r
B_r-r^2A^2C^2EF^2B_\theta C_\theta -r^2ABC^2EF^2A_\theta C_\theta
\nonu
\A \A +r^2A^2C^3F^2B_\theta E_\theta-r^2A^2CE^2F^2B_r
E_r-r^2ABCE^2F^2A_r E_r-6AB^2CE^3A_r B_r \nonu
 \A \A+2rABCE^3F^2A_r+2r^2A^2CE^3FB_r F_r+2r^2ABCE^3FA_r
F_r+2r^2A^2C^3EFB_\theta F_\theta-6AB^2C^3EA_\theta B_\theta \nonu
\A \A +2r^2ABC^3EFA_\theta
F_\theta-3A^2BC^3E{B_\theta}^2-3A^2BCE^3{B_r}^2-3B^3C^3E{A_\theta}^2+
2r^2AC^3EF^2A_\theta B_\theta \nonu
\A \A
+4r^2BC^3EF^2{A_\theta}^2-3B^3CE^3{A_r}^2-2r^2A^2BC^3FE_\theta
F_\theta+2r^2A^2BCE^2FE_r F_r+2rA^2BCE^2F^2E_r \nonu
\A \A -2rA^2BE^3F^2C_r-2r^2A^2BE^3FC_r F_r+2r^2A^2BC^2EFC_\theta
F_\theta-2r^2A^2BC^2F^2C_\theta E_\theta+4rA^2BCE^3F F_r \nonu
\A \A -2r^2A^2BE^2F^2C_r E_r-rA^2C^2E^3FB_r-rABC^2E^3FA_r
\Biggr)=0, \ea \ba \A \A {1 \over 2r^2A^4C^3E^3F^2}
\Biggl(2r^2A^2CE^2F^2 E_{rr}-2r^2A^2E^3FC_r
F_r+4rA^2CE^3FF_r+2rA^2CE^2F^2E_r \nonu
\A \A +2r^2A^2CE^2FE_r F_r+2r^2A^2C^3EF F_{\theta
\theta}-2r^2A^2C^3FE_\theta F_\theta+2r^2A^2C^2EFC_\theta
F_\theta+2r^2A^2CE^3FF_{rr}\nonu
\A \A -2r^2A^2C^2F^2C_\theta E_\theta-2r^2A^2E^2F^2C_r
E_r+2r^2A^2C^2EF^2C_{\theta \theta}-6ABC^3EA_\theta
B_\theta-2rA^2E^3F^2C_r \nonu
\A \A -6ABCE^3A_r
B_r-3A^2C^3E{B_\theta}^2-3B^2EC^3{A_\theta}^2-3A^2CE^3{B_r}^2-3B^2CE^3{A_r}^2
\Biggr)=0, \ea \ba \A \A {1 \over 2rA^3C^3E^3F} \Biggl(
2rBC^3EF{A_\theta}^2-rA^2C^3EFB_{\theta \theta}+rA^2C^3FB_\theta
E_\theta+2rBCE^3F{A_r}^2-rA^2CE^3FB_{rr} \nonu
\A \A+rABC^3FA_\theta E_\theta+rA^2E^3FB_r
C_r-rABCE^3FA_{rr}+rABE^3FA_r C_r+A^2C^2E^3B_r+ABC^2E^3A_r \nonu
\A \A -rABC^3EFA_{\theta \theta}-rA^2CE^2FB_r E_r-rABC^2EFA_\theta
C_\theta-rABCE^2FA_r E_r\nonu
\A \A - rA^2C^2EFB_\theta C_\theta \Biggr)=0, \ea where
$A_{r}=dA/dr$, $A_{rr}=d^2A/dr^2$, $A_{\theta}=dA/d\theta$
$A_{\theta \theta}=d^2A/d\theta^2$.

A special vacuum solutions of equations (19)$\sim$(25) are \vspace{.3cm}\\
I) The solution in which the five functions $A$, $B$, $D$, $F$ and
$H$ take the form \ba A \A=\A {1\over \sqrt{1-\displaystyle{a_1
\over r}}},\nonu
B \A=\A 0, \nonu
C \A=\A \displaystyle \sqrt{{1 \over {1-\displaystyle{a_1 \over
r}}}},\nonu
E \A =\A r, \nonu
F \A =\A \sin\theta,
 \ea
 where $a_1$ is a constant of integration.
 Using (26) in (18), the metric tensor takes the form
 \be ds^2= -\eta_1 dt^2 +{dr^2 \over \eta_1} +r^2d\Omega^2, \ee
where \be \eta_1(r)=(1-{a_1 \over r}). \ee which is the
Schwarzschild solution when the constant $a_1=2m$.

II) The solution in which the five functions $A$, $B$, $D$, $F$
and $H$ take the form
 \ba
 A \A=\A {1\over
\sqrt{1-\displaystyle{a \rho \over \Sigma}}},\nonu
B \A=\A {-a h \rho \sin^2\theta \over \sqrt{\Sigma
(\rho^2+h^2\cos^2\theta-a \rho)}}, \nonu
C \A=\A \sqrt{{\Sigma \over \Delta}},\nonu
E \A =\A \sqrt{\Sigma}, \nonu
F \A =\A \sqrt{(\rho^2+h^2){\sin^2\theta \over \rho^2}+{a h^2 \rho
\sin^4\theta \over \rho^2 \Sigma}+{a^2h^2 \sin^4\theta \over
\Sigma(\rho^2+h^2 \cos^2\theta-a \rho)}},
 \ea
 where $a$, $h$ are two constants of integration and $\Sigma$, $\Delta$ are given by
 \be
 \Sigma=\rho^2+h^2\cos^2\theta, \qquad \Delta=\rho^2+h^2-a \rho.
 \ee
 The metric tensor associated with the tetrad (18) in this case is
 given by
 \be
ds^2= -({1-\displaystyle{a \rho \over \Sigma}}) dt^2 +{\Sigma
\over \Delta} d\rho^2 +\Sigma
d\theta^2+\left\{(\rho^2+h^2)\sin^2\theta+{a\rho h^2\sin^4\theta
\over \Sigma} \right \}d\phi^2+2{a\rho h \sin^2\theta \over
\Sigma} dt d\phi, \ee which is the Kerr metric written in the
Boyer-Lindquist coordinates. Using the solution (29) the tetrad
(18) takes the form \be \left( {b^i}_\mu \right)= \left( \matrix{
\displaystyle i \sqrt{1-\displaystyle{a \rho \over \Sigma}} & 0 &
0 & -i \displaystyle{a h \rho \sin^2\theta \over \sqrt{\Sigma
(\rho^2+h^2\cos^2\theta-a \rho)}} \vspace{3mm} \cr 0 &
\displaystyle \sqrt{{\Sigma \over \Delta}}\cos\phi & 0 &
-\rho\sqrt{(\rho^2+h^2){\sin^2\theta \over \rho^2}+\displaystyle
{a h^2 \rho \sin^4\theta \over \rho^2 \Sigma}+\displaystyle
{a^2h^2 \sin^4\theta \over \Sigma (\rho^2+h^2 \cos^2\theta-a
\rho)}} \sin\phi \vspace{3mm} \cr 0 & 0 & \sqrt{\Sigma} & 0
\vspace{3mm} \cr 0 & \displaystyle \sqrt{{\Sigma \over \Delta}}
\sin\phi & 0 & \rho \sqrt{(\rho^2+h^2) \displaystyle {\sin^2\theta
\over \rho^2}+\displaystyle {a h^2 \rho \sin^4\theta \over \rho^2
\Sigma}+\displaystyle {a^2h^2 \sin^4\theta \over \Sigma
(\rho^2+h^2 \cos^2\theta-a \rho)}} \cos\phi \cr } \right). \ee

 Thus a vacuum solution which gives the Kerr metric has been
given. The parallel vector fields (32) are axially symmetric in
the sense that they are form invariant under the transformation
\ba \A \A \bar{\phi}\rightarrow \phi+\delta \phi, \qquad
\bar{b^0}_0 \rightarrow {b^0}_0, \qquad  \bar{b^1}_1 \rightarrow
{b^1}_1 \cos\delta \phi+{b^3}_3 \sin\delta \phi,\nonu
\A \A \bar{b^2}_2 \rightarrow {b^2}_2, \qquad \bar{b^3}_3
\rightarrow {b^3}_3 \cos\delta \phi-{b^1}_1 \sin \delta \phi. \ea

The solution (26) satisfy the field equation (7) and (8), but the
solution (32) is a solution to the field equation (7) only, i.e.,
the solution (32) is a solution of general relativity. It is of
interest to note that general relativity has a solution which give
the Kerr metric in which the parallel vector fields take the form
\be \left( {b^i}_\mu \right)_{Sq}= \left( \matrix{ \displaystyle i
\sqrt{1-\displaystyle{a \rho \over \Sigma}} & 0 & 0 & -i
\displaystyle{a h \rho \sin^2\theta \over \sqrt{\Sigma
(\rho^2+h^2\cos^2\theta-a \rho)}} \vspace{3mm} \cr 0 &
\displaystyle \sqrt{{\Sigma \over \Delta}} & 0 & 0
 \vspace{3mm} \cr 0 & 0 & \sqrt{\Sigma} & 0
\vspace{3mm} \cr 0 & 0& 0 & \rho \sqrt{(\rho^2+h^2) \displaystyle
{\sin^2\theta \over \rho^2}+\displaystyle {a h^2 \rho \sin^4\theta
\over \rho^2 \Sigma}+\displaystyle {a^2h^2 \sin^4\theta \over
\Sigma (\rho^2+h^2 \cos^2\theta-a \rho)}} \cr } \right), \ee where
$\Sigma$ and $\Delta$ are given by (30). As is clear from (34)
that this is just the square root of the Kerr metric.

\newsection{The Physical Meaning of $a$ and $h$}
Following Toma \cite{Tn} to clarify the physical meaning of the
constants $a$ and $h$ in the solution (32), we consider the weak
field approximation \cite{SN} \be {b^k}_{\mu}(x)=
{\delta^k}_{\mu}+{a^k}_{\mu}, \quad |{a^k}_{\mu}|<<1, \ee the
field $a_{\mu \nu}$ can be expressed as  \be a_{\mu \nu}={1 \over
2}h_{\mu \nu}+A_{\mu \nu}, \ee with $h_{\mu \nu}=h_{\nu \mu}$ and
$A_{\mu \nu}=-A_{\nu \mu}$. The gravitational field equations (7)
and (8) takes the forms \ba \Box{\overline h}_{\mu \nu} \A=\A
-2\kappa T_{(\mu \nu)},\nonu
 \Box A_{\mu \nu} \A=\A -\lambda
T_{[\mu \nu]}. \ea Here the d'Alembertian operator is given by
$\Box = \partial^{\mu} \partial_{\mu}$,  ${\overline h}_{\mu \nu}$
denotes \be {\overline h}_{\mu \nu}=h_{\mu \nu}-{1 \over
2}\eta_{\mu \nu}h, \qquad h=\eta_{\mu \nu}h^{\mu \nu}. \ee

Since we have solve the field equations (7) and (8) in the vacuum
case then from (37) we require \be A_{\mu \nu}=0.\ee The tensor
$T_{[\mu \nu]}$ is related to the spin density \be {S_{k
l}}^\mu=i{\partial {\cal L}_m \over \partial \psi_{,\mu}} S_{kl}
\psi \ee of the source field $\psi$ through \be \sqrt{-g} T_{[\mu
\nu]}={1 \over 2}{b^k}_\mu {b^i}_\nu \partial_\lambda
{S_{kl}}^{\lambda}. \ee
 Here $S_{kl}$ is the representation group of the Lie algebra of
 the Lorentz group to which $\psi$ belongs. Using (39) in (37) we
 get
 \be
 \partial_\mu{ S_{kl}}^\mu=0.\ee
 Using the condition (42), the physical meaning of $a$ and $h$ are
 given by
 \ba
 a \A=\A {\kappa M \over 4 \pi}\nonu
 h \A= \A -{J \over M},
 \ea
 where M is the gravitational mass of a central gravitating body
 and J represent the angular momentum of the rotating source. The
 relation (43) is obtained by comparing the metric (31) with the
 metric \cite{Tn,LT}
 \be
 ds^2=-\left(1-{\kappa M \over 4r \pi} \right)dt^2+\left(1+{\kappa M \over 4
 r\pi}\right) dx^a dx^a-{\kappa J \over 2r \pi} \sin^2\theta dt
 d\phi.
 \ee

\newsection{Study of Singularities}
In teleparallel theory of gravity by singularity of space-time we
mean the singularity of the scalar concomitants  of the torsion
and curvature tensors.

The space-time given by (32) does not have singularities for the
Ricci scalar and for the concomitants of the Ricci tensor because
we have  \be R=R^{\mu \nu}R_{\mu \nu}=0,\ee using (10) and (32) we
obtain \be R^{\mu \nu \rho \sigma}R_{\mu \nu \rho \sigma}= -12 a^2
{\left (h^6 \cos\theta^6-15\rho^2 h^4\cos\theta^4+15 \rho^4
h^2\cos^2\theta-\rho^6 \right) \over (\rho^2+h^2\cos^2\theta)^6 }.
\ee As is clear from (46) that there is a singularity at $(\rho, \
\theta)=(0, \ \pi/2)$ when $h \neq 0$, which agrees with the
singularity of solution (34).
 The scalar of the torsion tensor and the
irreducible components of it \cite{HS} are given by \ba T^{\mu \nu
\lambda} T_{\mu \nu \lambda} \A =\A  {F_1(\rho,\theta) \over
\sin^{3/2}\theta(\rho^2+h^2-a\rho)(\rho^2+h^2\cos^2\theta-a\rho)^{5/2}(\rho^2+h^2\cos^2\theta)^3
},\nonu
t^{\mu \nu \lambda}t_{\mu \nu \lambda} \A=\A {F_2(\rho,\theta)
\over
\sin^{3/2}\theta(\rho^2+h^2-a\rho)(\rho^2+h^2\cos^2\theta-a\rho)^{5/2}(\rho^2+h^2\cos^2\theta)^3
},\nonu
V^\mu V_\mu \A =\A {F_3(\rho,\theta) \over
\sin^2\theta(\rho^2+h^2-a\rho)(\rho^2+h^2\cos^2\theta)^3 },\nonu
a^\mu a_\mu \A=\A {F_4(\rho,\theta) \over
(\rho^2+h^2-a\rho)^3(\rho^2+h^2\cos^2\theta)^5},
 \ea
 where $F_1(\rho,\theta)$, $F_2(\rho,\theta)$, $F_3(\rho,\theta)$
 and $F_4(\rho,\theta)$ are some complicated functions in $\rho$
 and $\theta$.
From (47) we see that,  when $h \neq 0$, there are singularities
of the scalar of the torsion tensor, the scalar of the traceless
part and the scalar of the basic vector  given by $\theta=0$
and/or $(\rho, \ \theta)=(0, \ \pi/2)$, but for the axial vector
the singularity is given by $(\rho,\ \theta)=(0,\ \pi/2)$. Also
all the above scalars have a common singularity at $\rho=0$ when
$h=0$
\newsection{Main results}

The results of the proceeding sections can be summarized as
follow:\vspace{.6cm}\\ 1) An exact  solution (26) for the field
equations (7) and (8) which gives the Schwarzschild metric has
been obtained.\vspace{.4cm}\\ 2) An exact solution (32) for the
field equation (7) which gives the static Kerr metric is obtained.
It is axially symmetric, but differs from (34) of general
relativity. There is a modification of the Kerr metric for a
source that carries an electric charge e. Replacing the
 definition of $\Delta$ given in (30) by
 \[\Delta=\rho^2+h^2-a \rho+e^2,\] leads to the Kerr-Newman metric
 \cite{Bo}. It is shown that special relativity for the spin
 $\displaystyle{1 \over 2}$ electron can be seen to emerge from
 Kerr-Newman metric \cite{Bg}. Also it is shown that \cite{Bg} a particle
 can be treated as a relativistic vortex, that is a vortex where the
 velocity of a circulation equals that of light or a spherical shell,
 whose constituents are again rotating with the velocity of light
 or as a black hole described by the Kerr-Newman metric for a spin
 $\displaystyle{1 \over 2}$  particles. \vspace{.4cm}\\
3) The space-time given by (32) has a singularity at $(\rho, \
\theta)=(0, \ \pi/2)$ for the concomitants of the Riemann
Christoffel tensor and the concomitants of the axial vector part
when $h\neq 0$ but for the concomitants of the torsion, tracless
part and the basic vector there are singularities at $(\rho, \
\theta)=(0, \ \pi/2)$ and/or $\theta=0$ when $h\neq 0$ and all the
concomitants have a common singularity at $\rho=0$ and $h=0$.\vspace{.4cm}\\
4) There are two solutions of general relativity both of them gave
the Kerr metric, but the parallel vector fields reproducing them
are quite different in its structures in spite that the
singularities of them are quite the same.  We believe that the
physical contents of these two parallel vector fields (32) and
(34) are quite different and this will be our future work by
studying the energy contents for the both tetrad.\vspace{.4cm}\\
5) We have solved the non linear partial differential equations
(19)$\sim$(25) and obtained two different exact vacuum solutions.
From those partial differential equation one can get many physical
solutions and may be can solve them in the general case. This also
will be our future work.

\end{document}